\documentclass[10pt,conference]{IEEEtran}

\usepackage[utf8]{inputenc}
\usepackage{url}
\usepackage{enumitem}
\usepackage{flushend}

\ifCLASSINFOpdf
   \usepackage[pdftex]{graphicx}
\else
\fi

\usepackage{amssymb}
\usepackage{xspace}
\newcommand{\etal}{et al.\xspace}
\newcommand{\ie}{i.e.,\xspace}

\newcommand{\fig}[1]{Fig.~\ref{#1}}

\usepackage[disable]{todonotes}
\newcommand{\mika}[1]{\todo[color=blue!40, inline]{\footnotesize{Mika: #1}}}

\newcommand{\mael}[1]{\todo[color=red!40, inline]{\footnotesize{Maël: #1}}}

\newcommand{\rev}[2]{\todo[color=pink, inline]{\footnotesize{Reviewer #1: #2}}}

\begin{document}
%
\title{Abnormal Working Hours:  Effect of Rapid Releases and Implications to Work Content}

\author{
\IEEEauthorblockN{Maëlick Claes\IEEEauthorrefmark{1}, Mika Mäntylä\IEEEauthorrefmark{1}, Miikka Kuutila\IEEEauthorrefmark{1} and Bram Adams\IEEEauthorrefmark{2}
}
\IEEEauthorblockA{\IEEEauthorrefmark{1} M3S, ITEE, University of Oulu, Finland \\
Email: firstname.lastname@oulu.fi}
\IEEEauthorblockA{\IEEEauthorrefmark{2} MCIS, Polytechnique Montreal, Canada\\
Email: bram.adams@polymtl.ca}
}

%


\maketitle

%
%

\begin{abstract}

During the past years, overload at work leading to psychological
diseases, such as burnouts, have drawn more public attention. This
paper is a preliminary step toward an analysis of the work patterns
and possible indicators of overload and time pressure on software
developers with mining software repositories approach. We explore the
working pattern of developers in the context of Mozilla Firefox, a
large and long-lived open source project. To that end we investigate
the impact of the move from traditional to rapid release cycle on work
pattern. Moreover we compare Mozilla Firefox work pattern with another
Mozilla product, Firefox OS, which has a different release cycle than
Firefox. We find that both projects exhibit healthy working patterns,
i.e. lower activity during the weekends and outside of office
hours. Firefox experiences proportionally more activity on weekends
than Firefox OS (Cohen's $d = 0.94$). We find that switching to rapid
releases has reduced weekend work (Cohen's $d = 1.43$) and working
during the night (Cohen's $d = 0.45$). This result holds even when we
limit the analyzes on the hired resources, i.e. considering only
individuals with Mozilla foundation email address, although, the
effect sizes are smaller for weekends (Cohen's $d = 0.64$) and nights
(Cohen's $d = 0.23$). Moreover, we use dissimilarity word clouds and
find that work during the weekend is more technical while work during
the week expresses more positive sentiment with words like “good” and
“nice”. Our results suggest that moving to rapid releases have
positive impact on the work health and work-life-balance of software
engineers. However, caution is needed as our results are based on a
limited set of quantitative data from a single organization.

\end{abstract}



%
\IEEEpeerreviewmaketitle


\section{Introduction}
Factors relating to occupational health and wellbeing are seldom
addressed in the software engineering body of knowledge. This is
surprising considering the productivity in software development is
largely related to peopleware, and large productivity differences have been
documented~\cite{mcconnell2011does}. Furthermore, replacing key
individuals with in-depth knowledge of the product with new employees
may take years as the new employees need ample time to gain knowledge
about the product. As the matters of occupational health and wellbeing
can reduce staff turnover, sick days and decrease mortality among
software engineers, addressing them with mining software repositories
becomes an important objective.

Long working hours have been associated with depressive state,
anxiety, sleep condition, and coronary heart disease
~\cite{bannai2014association}. Importance of weekend recovery has been
showed to affect weekly job performance personal initiative,
organizational citizenship behaviour, and lower perceived effort
~\cite{binnewies2010recovery}. Another study showed that psychological
detachment during off-work time reduced emotional exhaustion caused
by high job demands and helped to maintain work engagement
~\cite{sonnentag2010staying}. Given the importance of the off-work
time in employee recovery we investigate software developers work
patterns from large open source project Mozilla Firefox with many
hired resources. We are particularly interested in work performed
outside of office hours as such working hours can acts as a proxy for
job related stress and time pressure conditions and suggest non-sufficient
detachment from work.

Mozilla Firefox is a web browser with widespread and long lived
history. Firefox makes an interesting case study as it is supported by
the Mozilla Foundation, that pay software engineers to develop
it. Moreover, it moved from a slow to a fast release cycle in
2011. Until the release of Firefox 4.0 in March 2011, new major
versions were released not more than once in a year and parallel
versions kept being maintained for security updates. Following Firefox
4.0, a new major version is to be released every 6 weeks and supersede
its previous version. This change from a slow to a fast release cycle
has a potential impact on work patterns.

We investigate the following research questions:
\begin{description}
\item[RQ1] What is the developers' usual pattern of work activity? We
  look when developers are more likely to be active in both Mozilla
  Firefox and Firefox OS.
\item[RQ2] Is there a difference in the developers' work pattern when
  switching to rapid releases? We compare the difference in work
  pattern in Mozilla Firefox before and after switching to rapid
  releases.
\item[RQ3] Do developers show a different behavior outside office
  hours? Using basic text mining techniques we analyze the vocabulary
  more likely to be used outside and during office hours.
\end{description}

\rev{2}{It would be useful to define what a work pattern is early in the paper}
\rev{2}{RQ3: Do developers show a different behavior outside office
  hours? $->$ when I first read this, I was confused as to what you mean
  by "different behavior". Whad kind of behavior? It only becomes
  clearer later in the paper.}

This paper is structured as follow. In
Section~\ref{sec:related} we list related work. In
Section~\ref{sec:data} we give details on how we extracted data to
answer our research questions that we answer in
Section~\ref{sec:analysis}. We then present the threats to validity
that can impact our results in Section~\ref{sec:threats} and conclude
in Section~\ref{sec:conclusion}.



\section{Related Work}\label{sec:related}

Sall \etal~\cite{Sall2007} studied weekend work activity patterns in
San Francisco Bay Area using surveys. Results indicate that a host of
variables affect the likeliness of working during the weekend. They
found that gender, race, type of work and income affect the work
pattern. Individuals are more likely to work out of home during
weekends in winter season than other seasons.

Wang \etal~\cite{DBLP:journals/corr/abs-1208-2686} examined work
patterns of scientists by looking at the amount of scientific papers
being downloaded in different days. Scientists work for 60-70\% as
much during the weekend as during the week. Time worked during
weekends differs by country: scientists work proportionally more
during weekends in China than in USA and Germany.

Binnewies \etal~\cite{binnewies2010recovery} investigated the importance of
recovery during weekend and its implications on work performance. Data
from surveys indicate that experiences of psychological detachment,
relaxation and mastery during weekend were positively correlated with
of being recovered beginning the working week, which in turn was
positively related to self reported work performance.

McKee~\cite{McKeei2750} collected the findings of increased mortality
rate in hospitals during the weekend, with explanations ranging from
more seriously sick patients to less experiences staff. At extreme,
the weekend effect has been observed to be 44\% higher odds of
mortality in Friday compared to Monday \cite{Aylinh4652}. However,
multiple sources state conflicting evidence on the source of this
effect \cite{McKeei2750} ,\cite{Aylinh4652}, \cite{Alexander2010}.

In software context, it has been observed that commits made between
mid-night and 4 AM contain more bugs, while commits made between 7 AM
and noon contain the least \cite{Eyolfson2011}. However, sentiment
analysis of commit messages finds more negative emotion on Mondays
compared to other days \cite{Guzman2014}.

Khomh \etal~\cite{Khomh2015} studied the impact of Firefox's fast
release cycle on post-release bugs. They found that not only the new
release cycle didn't increase the number of bugs, but bugs
are also fixed faster. The real challenges faced by the switch of
release cycle was related to the need of automating the release
engineering process.



\section{Data extraction}\label{sec:data}

According to Mozilla's guidelines, every change must be reviewed by
its component owner. For that to happen, an entry must be added in the
bug tracker regardless of the type of change (bug/security fix,
performance enhancements, new feature, etc). We extracted bug comments
from Mozilla's Bugzilla
repository\footnote{\url{http://bugzilla.mozilla.com}} using the
GrimoireLab \mika{often academic tool makers ask for proper citations. }\mael{AFAIK they don't have a paper on these tools}tools\footnote{\url{https://grimoirelab.github.io/}}. We
filtered the bugs to only keep the ones related to the following
\emph{products}: \emph{Firefox}, \emph{Core} and \emph{Firefox OS}.

\rev{1}{In this paper, the authors only consider two projects, i.e.,
  Firefox and Firefox OS. First, I do not understand what is the
  difference between Firefox and Firefox OS, since they look
  similar. Also, whether Firefox OS has a rapid release too?}

\mael{Modified the abstract to take this into account. Is it enough?}
\mika{I added the usual we must be cautious sentence to cap off the abstract}

\rev{1}{It would be interesting to investigate more projects from more
  open source communities (e.g., Eclipse, Apache, and OpenOffice). As
  a short paper, it is OK to only investigate one or two projects. But
  when the authors plan to extend it to be a full conference paper or
  a journal paper, they should include more projects.}

However Bugzilla's data lacks time zone information. In order to be
able to identify when people are working, we extracted timezone
information from Mozilla unified Mercurial
repository\footnote{\url{https://hg.mozilla.org/mozilla-unified}}. We
matched contributors to the bug tracker with contributors with the
source code repository using e-mail addresses. This allows us to focus
on bug tracker contributors who have also contributed code, and thus
can be considered as developers. While we were able to associate a
timezone to only 2637 out of 145691 (1.8\%) of the bug tracker
contributors, 5.86 millions of comments out 11.11 millions (53\%) were
authored by them.

Each bug can have files, which are often patches, attached to
it. Because we want to focus on development activity, we will only
consider comments that were related to patch submission and code
review. These comments can easily be identified as the comments
starting with ``Created attachment'' and ``Comment on
attachment''. However we ignore those starting lines when analyzing
the text of the comments. We also ignored comments that were posted by
e-mail addresses known to be used by bots.

Analyses comparing Firefox and Firefox OS were made using
comments written between 2012-12-21 and 2016-01-03. These dates
correspond to the \emph{feature complete} date of Firefox OS 1.0 and
\emph{code complete} release date of Firefox OS 2.5. In other words, we limit our analysis to the time when professional development of Firefox OS took place. Firefox browser development was professional during the entire time.

\rev{1}{Why the authors use number of bug comments but not the commits
  submitted to version control system to study the work patterns? It
  would be interesting to investigate whether the conclusions are
  still the same if consider the number of commits submitted to VCS.}


\section{Empirical analysis}\label{sec:analysis}

\subsection{What is the developers' usual work pattern?}

\begin{figure}[!htpb]
  \centering
  \includegraphics[width=0.7\columnwidth]{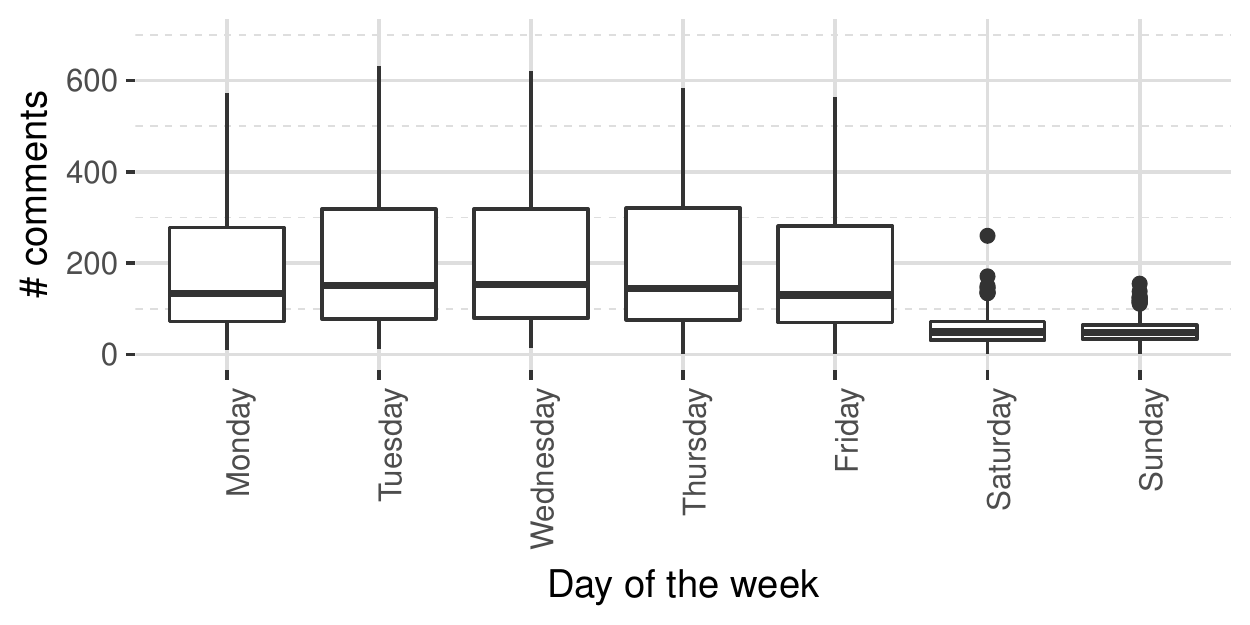}
  \caption{Distribution of the number of comments for Firefox
    depending on the day of the week.}
  \label{fig:firefox-weekdays}
\end{figure}

First we computed the number of comments related to attached files for
each day. 
\fig{fig:firefox-weekdays}
shows that comments are not evenly distributed over the days of the
week. Indeed fewer comments are posted during the weekend.

\begin{figure}[!htpb]
  \centering
  \includegraphics[width=0.7\columnwidth]{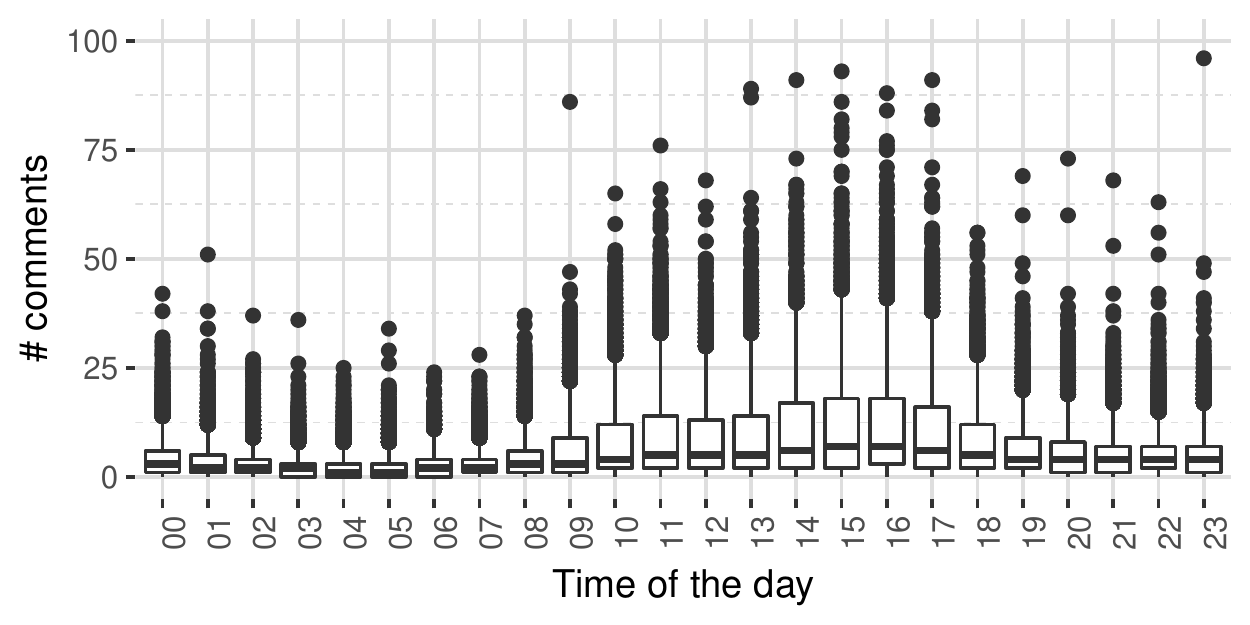}
  \caption{Distribution of the number of comments for Firefox
    depending on the time of the day.}
  \label{fig:firefox-hours}
\end{figure}

Similarly, we computed the number of comments for each hour of the
day. \fig{fig:firefox-hours} shows that most activity occurs between 9
am and 7 pm. Overall the usual work pattern of Firefox developers
follow regular office hours with low activity during the night and
during the weekends.

\begin{figure}[!htpb]
  \centering
  \includegraphics[width=0.7\columnwidth]{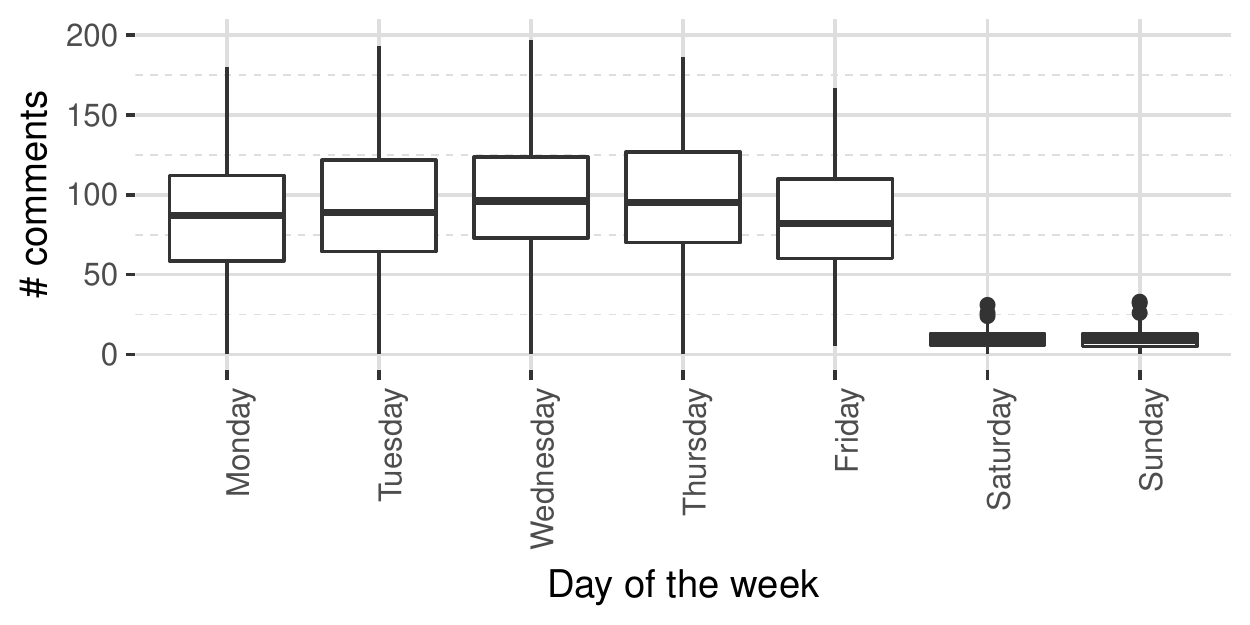}
  \caption{Distributions of the number of comments for Firefox OS by day of the week.}
  \label{fig:firefoxos-weekdays}
\end{figure}

\begin{figure}[!htpb]
  \centering
  \includegraphics[width=0.7\columnwidth]{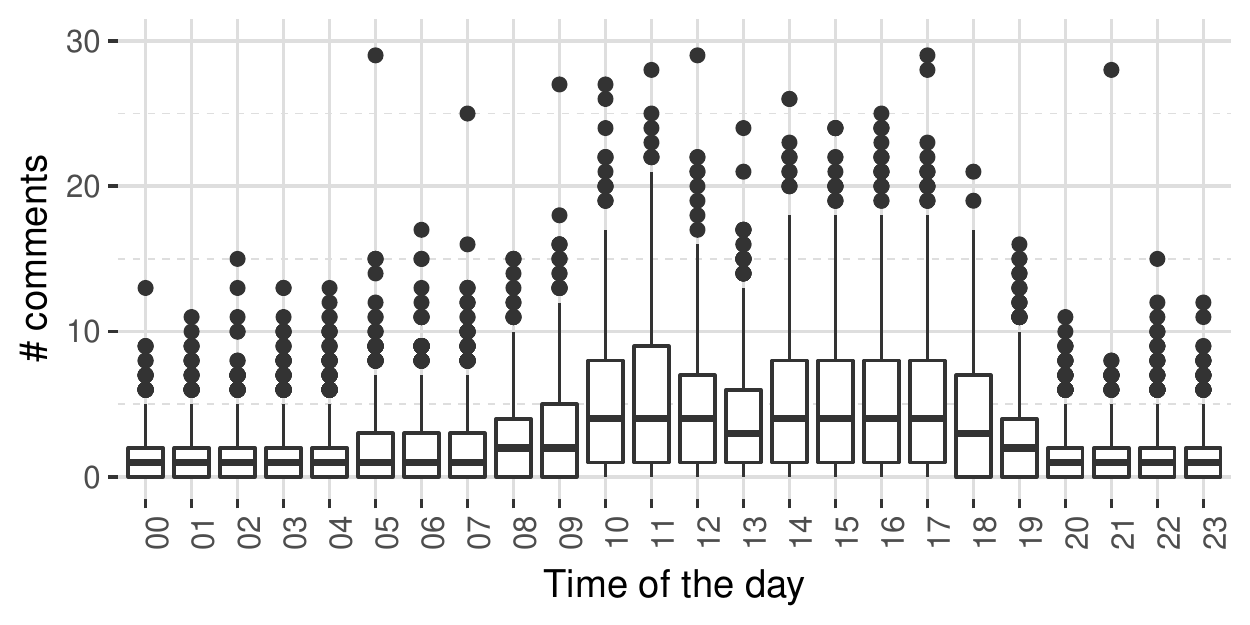}
  \caption{Distributions of the number of comments for Firefox OS by time of the day.}
  \label{fig:firefoxos-hours}
\end{figure}

We compare Firefox work pattern with another Mozilla project: Firefox
OS. \fig{fig:firefoxos-weekdays} and \fig{fig:firefoxos-hours} shows
that Firefox OS developers follow the same work pattern.
Based from these observations, we identified the period from 8 am to 7
pm as the day and from 8 pm to 7 am as the night. For both systems we
computed the proportion of daily comments made during the night, and
the proportion of comments made during a weekend day compared to a
weekday. 
We observed that relative activity in Firefox OS is significantly
lower during weekends than for Firefox. While we both observe a
statistically significant difference between projects for both weekend
and night work, the effect size of the weekend work is large (Mann–Whitney U test $p$-value
$< 0.001$, Cohen's $d = 0.94$, Cliff's $\delta = 0.77$) and neglible for night work
(Mann–Whitney U test $p$-value $< 0.01$, Cohen's $d = 0.06$, Cliff's
$\delta = 0.06$). On average
for Firefox, a weekend day receives 24\% of the comments a weekday
receives, while for Firefox OS a weekend day receives 12.2\% of the
comments of a weekday.

This means that even though Firefox shows a rather healthy work
pattern, differences can exist between projects. One possible
explanation is the difference in contributor profiles. Open source
projects attract hobbyist developers who might have a regular paid job
and contribute to the project during their free time. Differences in
the amount of such developers can potentially explain differences in
work pattern. While we observe lower weekend and night activity when
focusing solely on contributors using a Mozilla e-mail address, there
is still an important difference for the weekend activity between
Firefox and Firefox OS (Mann–Whitney U test $p$-value $< 0.001$,
Cohen's $d = 0.53$, Cliff's $\delta = 0.6$).

\rev{2}{Although interesting, I wonder if this paragraph is really
  necessary. I would suggest to cut it off if you need space.}

\mael{Remove paragraph above if not enough space}





\subsection{Is there a difference in the developers' work pattern when
  switching to rapid releases?}

\begin{figure*}[!htpb]
  \centering
  \includegraphics[width=\columnwidth]{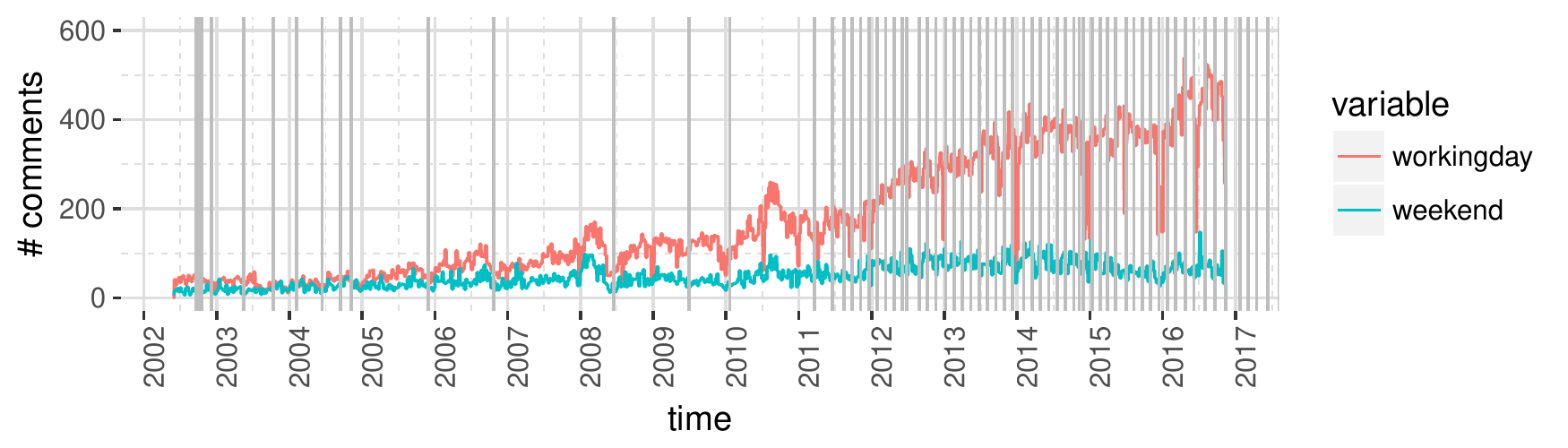}
  \includegraphics[width=\columnwidth]{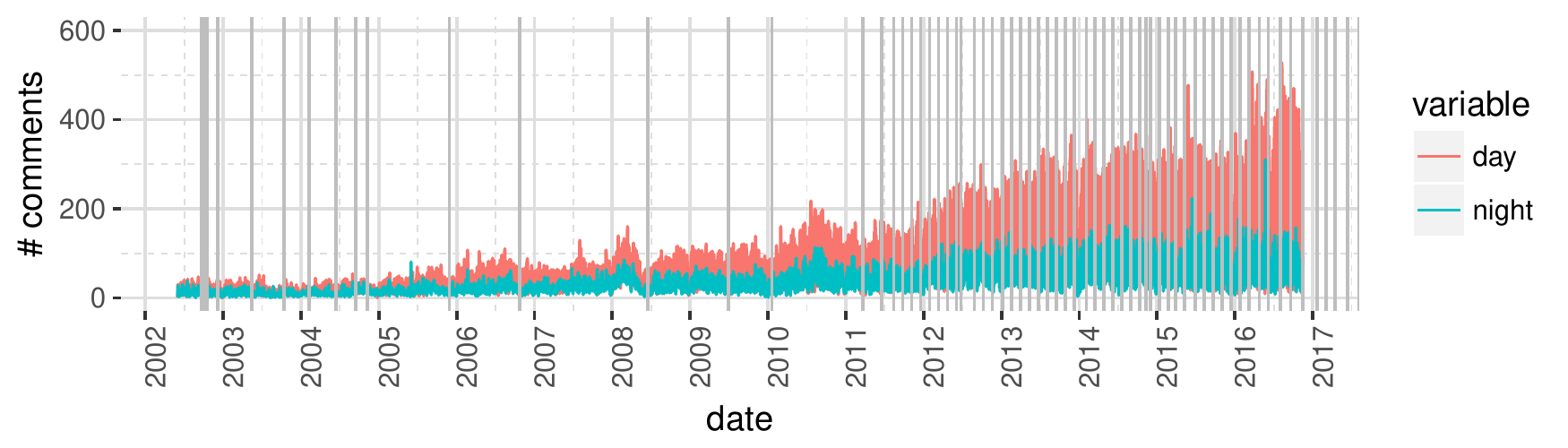}
  \caption{Evolution of the number of comments for Firefox. Vertical gray
    lines represent major Firefox releases.}
  \label{fig:firefox-evol}
\end{figure*}




\fig{fig:firefox-evol} shows both the evolution over time of the
number of weekly comments posted during weekends and weekdays, and the
evolution of those posted during the night and during the day. While
the number of comments posted during office hours has increased
significantly since 2012, the same trend cannot be observed for the
night and weekend comments. We compared the distribution of the
percentage of comments during weekends and nights before and after
moving to rapid releases.
There is a statistically significant decrease in the relative amount
of comments made outside office hours. However this decrease is more
important for weekend work (Mann–Whitney U test $p$-value $< 0.001$,
Cohen's $d = 1.43$, Cliff's $\delta = 0.78$) than night work
(Mann–Whitney U test $p$-value $< 0.001$, Cohen's $d = 0.45$, Cliff's
$\delta = 0.33$). On average since the release of Firefox 4, a weekend
day receives 24.4\% of the comments a weekday receives, while before
Firefox 4, a weekend day used to receive 48.6\% of the comments of a
weekday.


Again, the difference might be explained by an increase in the number
of developers paid by Mozilla. Indeed 42.5\% of all comments since the
fast release cycle have been written by people using a Mozilla e-mail
address, while only 16.7\% of comments were before. However, even when
taking into accounts only comments posted with a Mozilla e-mail
address, we still observe a large decrease in the activity during
weekends (Mann–Whitney U test $p$-value $< 0.001$, Cohen's $d = 0.64$,
Cliff's $\delta = 0.52$) but a small to negligible one during the
nights (Mann–Whitney U test $p$-value $< 0.01$, Cohen's $d = 0.23$,
Cliff's $\delta = 0.05$). For comments posted with a Mozilla e-mail
addresses, on average since the release of Firefox 4, a weekend day
receives 17.2\% of the comments a weekday receives, while before
Firefox 4, a weekend day used to receive 52.4\% of the comments of a
weekday.

\subsection{Do developers show a different behavior outside office hours?}

\begin{figure}[!htpb]
  \centering
  \includegraphics[width=0.42\columnwidth]{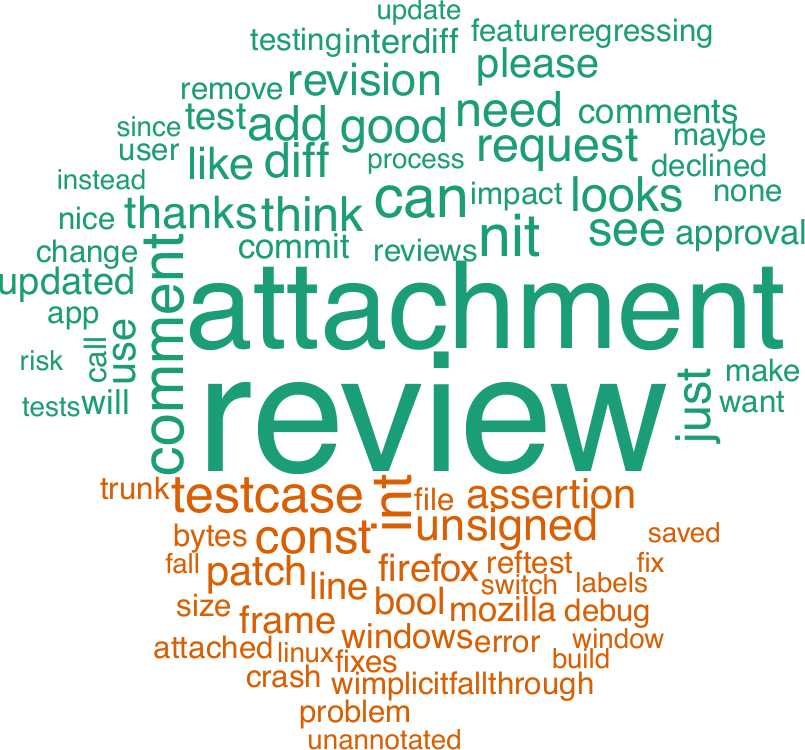}
  \caption{Comparison wordcloud of comments posted during weekends and
    weekdays. Words on the bottom, in orange, have a higher
    probability to appear in comments during weekends than words on
    the top, in green, which have a higher probability to appear in
    comments during weekdays.}
  \label{fig:wordcloud-weekends}
\end{figure}

In order to determine if there is a difference in the type of comments
developers write outside office hours, we computed the frequency of
appearance of the words from all comments. Using the \emph{wordcloud}
R
package\footnote{\url{https://cran.r-project.org/web/packages/wordcloud/index.html}}
we computed a comparison wordcloud depicted in
\fig{fig:wordcloud-weekends}. It highlights words that have a highest
probability to appear in a comment posted during a weekend or during a
weekday.

First many of the most likely words to appear in a weekend comment are
technical words. These include words like \emph{testcase},
\emph{assertion}, \emph{int}, \emph{bool}, \emph{file},
\emph{unsigned}, \emph{fix}, \emph{debug}, \emph{bytes}, \emph{error}
or \emph{crash}. On the other hand, weekday comments are more likely
to have non technical words like \emph{add}, \emph{since},
\emph{instead}, \emph{will}, \emph{use}, \emph{see}, \emph{looks},
\emph{need}, \emph{think}, \emph{maybe} or \emph{can}. Moreover, we
also see positive or polite words like \emph{good}, \emph{please},
\emph{thanks} and \emph{nice}. This hints that developers tend to talk
more during weekdays than during weekends where they focus more on
technical aspects. The differences between the day and night are
similar but not as strong (figure ommited due to space
  restrictions).


Then words related to reviews are also more likely during the
weekdays. In particular, \emph{nit}, a shortcut of nitpicking, is used
in code review by the Mozilla community to indicate less import
comments to take into account. This might indicate that developers
tend to be more direct and focus on what's the most important when
working outside the regular work pattern.



\section{Threats to Validity}\label{sec:threats}

Our analysis depends on data extracted from repositories available
from online sources. Errors or incompleteness in these data sources
may impact the result of our analysis. In particular we relied on time
zone information available from Mozilla's Mercurial source code
repository. The time zone from this repository might not be accurate
to the developer's actual position. Moreover, we only used e-mail
addresses to match identities from the Mercurial repository with the
authors of comments in the Bugzilla repository. We are aware of some
developers using a different e-mail address in each repository.

We limited our study to a single organization and the results obtained
might be specific to that organization culture and habits of its
developers. Similarly this can also explain similarities in work
pattern between Firefox and Firefox OS.

In order to identify developers hired by Mozilla, we used the domain
name of the e-mail address used as a proxy. We are aware of a few
major Mozilla developers who don't use a Mozilla e-mail
address. Finally, the scripts that we have developed for our empirical
analysis may still contain some bugs, and the obtained results may be
biased by some simplifying assumptions we have made during our
analysis.

\rev{3}{The issue about external validity is that IMHO it is very hard
  to study the effect of rapid releases by just observing one single
  datapoint with rapid release switch (Mozilla Firefox) compared to
  another single datapoint for which this did not happen (Firefox
  OS). But ok, if you properly confined the validity of your results,
  I would still be ok with that.}

\mael{Not sure how to take this one into account}
\mika{we could do interviews to triangulate firefox perssonel, use chat logs and other source of quantitative data or collect more projects. future work mainly}

\section{Conclusion and future work}\label{sec:conclusion}

In this paper we investigated software developers' work patterns and
possible indicators of overload and time pressure with mining software
repositories approach in the context of Mozilla Firefox. We find that
activity is lower outside office hours. We find a rather important
difference in the activity during weekend when comparing Firefox with
Firefox OS. We also find that the tendency to work outside office
hours, in particular during weekends, has reduced over the
years. While the differences can be explained by the amount of
developers hired by Mozilla, we find that the findings still hold when
limiting our analysis to Mozilla developers.

A common assumption of faster release cycle could propose that: as the
number of deadlines increases, this increases time pressure
related to deadlines, which increases abnormal work patterns,
\ie working during nights and weekends. However, it appears that the
opposite is true as switching to rapid releases has reduced the amount
of work performed outside office hours. Studies from psychology
partially support this: it was found that more frequent deadlines
reduces the likelihood of being late from the final
deadline~\cite{ariely2002procrastination}. Thus, it seems that rapid
release cycles have a positive effect on reducing work outside office
hours and lead to increased occupational health.

Moreover, we found that comments posted on weekends contained more
often technical terms than those posted on weekdays. On the other hand,
comments posted on weekdays usually contained more positive or polite
words, such as \emph{nice}, \emph{please} or \emph{thanks}. This hints
that developers tend to be more direct in their comments and focus
only on the technical aspect instead of formalities outside office
hours.

As future work, we intend to further study the impact of time pressure
and work patterns on software developers. We aim to compare multiple
projects to see whether we could empirically draw normative
recommendations of outside office work in software projects. We will
also use additional data sources, such as commits and chat logs, in
order to make our set of timestamp activity more complete, and conduct
interviews with development teams. Measuring the level of detachment
from work that is critical to recovery, should be studied and
similarly normative recommendation from empirical data should be
drawn. We also intend to study the impact of policies and guidelines
put in place by project managers, such as a fast release cycle, on
developers' activity and health. We also want to investigate further
the causes of outside office hours work by focusing on periods with
unusual high activity during the night or weekends.

\rev{3}{Concerning construct validity, I’m unsure to what extent the
  observation of code review timestamp would fully reflect development
  activity times. While developers can contribute to code reviews with
  patches and comments at specific times, they could have worked on
  the patches, or in general in various other development tasks, at
  times that you were not able to observe. Why commits were excluded
  from the analysis?}

\rev{2}{Finally, the results and conclusions suggest that the
  observable effects, i.e. changes in work patterns, are caused by a
  switch to a rapid release cycle. But the paper does not provide any
  detail on how this switch to a rapid release cycle occurred and
  which other changes in software process were made in the studied
  projects. As someone interested in software productivity and
  developer well-being, I'd be curious to know how that switch was
  implemented at Mozilla to better understand how their work practices
  changed with the introduction of rapid release cycles. I understand
  this may require another kind of study, which is out of the scope of
  this paper, but a brief explanation of Mozilla's process before and
  after RRC would be helpful.}

\rev{3}{The main concern I have here is how to explain the
  relationship between rapid releases and reduction of activity
  outside “regular” working hours and days. In order to properly
  understand such a relationship, a deeper analysis of a project’s
  organization is required. Intuitively, rapid releases would imply
  that developers tend to work more during office hours because they
  require to work closely together, do stand-up meetings,
  etc. However, is this really the case? did people really work close
  to each other, i.e., in the office and within the same time zone?
  Or, instead, the project involves geographically distributed
  developers? In order to better explain this, either you should
  interview developers, or analyze and discuss documents describing
  the typical projects’ work organization. This is why, as I said
  above, the purely quantitative observation is insufficient for this
  kind of study.}

\rev{3}{To better isolate confounding factors, I believe the best way
  to study this phenomenon is to observe the working pattern of a
  relatively small company (which team is mostly composed of
  co-located developers) during a switch towards rapid releases.}



\section*{Acknowledgments}

The authors have been supported by Academy of Finland grant 298020.





\bibliographystyle{IEEEtran}
\bibliography{biblio}

\end{document}